\documentclass[prd,preprintnumbers,twocolumn,amsmath,nofootinbib,amssymb]{revtex4}
\usepackage{graphicx,color,dcolumn,booktabs,bm}
\usepackage{longtable,lscape}
\usepackage{txfonts}
\usepackage{overpic}
\usepackage{amssymb}
\usepackage{epstopdf}
\usepackage{indentfirst}
\usepackage{feynmf}   %{feynmp}
\usepackage{slashed}  %for Feynman symbols
\usepackage{cases}
\usepackage{color}
\usepackage{float}
\usepackage{multirow}
\usepackage{ulem}
\usepackage{graphicx,color,dcolumn,booktabs,bm}
\usepackage{epsfig,dsfont,amssymb,amsmath,amsfonts,amsbsy,mathrsfs}
\graphicspath{{Figures/}} %

\usepackage{hyperref}
\hypersetup{colorlinks,citecolor=blue,anchorcolor=red,menucolor=red, linkcolor=red,filecolor=red,runcolor=red,urlcolor=blue,frenchlinks=red}

\makeatletter
\@addtoreset{equation}{section}
\makeatother

\allowdisplaybreaks

\begin{document}

\title{Radiative decays and magnetic moments of the predicted $B_c$-like molecules}

\author{Fu-Lai Wang$^{1,2,3,4,5}$}
\email{wangfl2016@lzu.edu.cn}
\author{Si-Qiang Luo$^{1,2,3,4,5}$}
\email{luosq15@lzu.edu.cn}
\author{Xiang Liu$^{1,2,3,4,5}$}
\email{xiangliu@lzu.edu.cn}
\affiliation{$^1$School of Physical Science and Technology, Lanzhou University, Lanzhou 730000, China\\
$^2$Lanzhou Center for Theoretical Physics, Key Laboratory of Theoretical Physics of Gansu Province, Lanzhou University, Lanzhou 730000, China\\
$^3$Key Laboratory of Quantum Theory and Applications of MoE, Lanzhou University,
Lanzhou 730000, China\\
$^4$Frontiers Science Center for Rare Isotopes, Lanzhou University, Lanzhou 730000, China\\
$^5$Research Center for Hadron and CSR Physics, Lanzhou University and Institute of Modern Physics of CAS, Lanzhou 730000, China}

\begin{abstract}
In this work, we first perform a systematic study of the transition magnetic moments and the corresponding radiative decay behaviors of the $B_c$-like molecular states associated with their mass spectra, where the constituent quark model is adopted by considering the $S$-$D$ wave mixing effect. Our numerical results show that the radiative decay properties can be considered as the effective physical observable to reflect the inner structures of these $B_c$-like molecular states. Meanwhile, we also discuss the magnetic moments of the $B_c$-like molecular states, and we find that the magnetic moment properties can be used to distinguish the $B_c$-like molecular states from the conventional $B_c$ mesonic states, which have the same quantum numbers and similar masses. We expect that the present study can inspire the interest of the experimentalist in exploring the electromagnetic properties of the $B_c$-like molecular states, especially the radiative decay properties.
\end{abstract}
\maketitle

\section{Introduction}\label{sec1}

At present, our understanding of the nonperturbative behavior of the strong interaction is still insufficient. The study of hadron spectroscopy is an important approach to deepen our understanding of the nonperturbative behavior of the strong  interaction, which has become an interesting research frontier in hadron physics. Since the discovery of the first charmoniumlike state $X(3872)$ by the Belle Collaboration in 2003 \cite{Choi:2003ue}, more and more new hadronic states have been reported in various high-energy physics experiments, bringing the study of hadron spectroscopy to a new stage
\cite{Liu:2013waa,Hosaka:2016pey,Chen:2016qju,Richard:2016eis,Lebed:2016hpi,Olsen:2017bmm,Guo:2017jvc,Brambilla:2019esw,Liu:2019zoy,Chen:2022asf,Meng:2022ozq,Amsler:2004ps,Swanson:2006st,Godfrey:2008nc,Yamaguchi:2019vea,Albuquerque:2018jkn,Yuan:2018inv,Ali:2017jda,Dong:2017gaw,Faccini:2012pj,Drenska:2010kg,Pakhlova:2010zza}.

In the past two decades, these novel phenomena have stimulated a broad interest among theorists in the study of the exotic hadronic states, such as multiquark states, hybrids, glueballs, and so on \cite{Liu:2013waa,Hosaka:2016pey,Chen:2016qju,Richard:2016eis,Lebed:2016hpi,Olsen:2017bmm,Guo:2017jvc,Brambilla:2019esw,Liu:2019zoy,Chen:2022asf,Meng:2022ozq,Amsler:2004ps,Swanson:2006st,Godfrey:2008nc,Yamaguchi:2019vea,Albuquerque:2018jkn,Yuan:2018inv,Ali:2017jda,Dong:2017gaw,Faccini:2012pj,Drenska:2010kg,Pakhlova:2010zza}. Especially, the observation of the molecular-type characteristic spectrum of the hidden-charm pentaquarks \cite{Wu:2010jy,Wang:2011rga,Yang:2011wz,Wu:2012md,Li:2014gra,Chen:2015loa,Karliner:2015ina,Chen:2015moa} in the $\Lambda_b \to J/\psi p K$ process \cite{Aaij:2019vzc}, which is represented by the $P_{c}(4312)$, $P_{c}(4440)$, and $P_{c}(4457)$ states, makes the hadronic molecular states attract more attention from the whole community \cite{Liu:2013waa,Hosaka:2016pey,Chen:2016qju,Richard:2016eis,Lebed:2016hpi,Olsen:2017bmm,Guo:2017jvc,Brambilla:2019esw,Liu:2019zoy,Chen:2022asf,Meng:2022ozq,Amsler:2004ps,Swanson:2006st,Godfrey:2008nc,
Yamaguchi:2019vea,Albuquerque:2018jkn,Yuan:2018inv,Ali:2017jda,Dong:2017gaw,Faccini:2012pj,Drenska:2010kg,Pakhlova:2010zza}.

Encouraged by this situation, the Lanzhou group focused on the $B_c$-like molecular states by checking the $S$-wave $D^{(*)}_{(s)}B^{(*)}_{(s)}$ interactions within the one-boson-exchange model \cite{Sun:2012sy}, and predicted the existence of a new type of the $B_c$-like molecular states. We are waiting for the experimental progress. Although some theoretical progress has been made on the $B_c$-like molecular states, our knowledge of the properties of the $B_c$-like molecules is not enough. There are some aspects around the $B_c$-like molecules that should be further explored.

We notice a phenomenon, i.e., some conventional $B_c$ mesonic states
\cite{Godfrey:1985xj,Eichten:1994gt,Gershtein:1994dxw,Zeng:1994vj,Ebert:2002pp,Godfrey:2004ya,Soni:2017wvy,Eichten:2019gig,Li:2019tbn,Li:2022bre} and some predicted $B_c$-like molecular states \cite{Sun:2012sy} have the same quantum numbers and similar masses, such as the $B_c(2P_1^{\prime})$ state and the $DB^*$ molecular state with $I(J^P)=0(1^+)$, the $B_c(2P_1)$ state and the $DB^*$ molecular  state with $I(J^P)=0(1^+)$, the $B_c(3P_1^{\prime})$ state and the $D_s^*B_s^*$ molecular state with $I(J^P)=0(1^+)$, the $B_c(3{}^3P_2)$ state and the $D_s^*B_s^*$ molecular state with $I(J^P)=0(2^+)$, and so on. How to distinguish them becomes a crucial problem. In this work, we propose to study their electromagnetic properties, by which answer whether or not the magnetic moment properties can be applied to distinguish the conventional $B_c$ mesonic states and the predicted $B_c$-like molecular states with the same quantum numbers and similar masses. Of course, we also study the transition magnetic moments and the corresponding radiative decay behaviors of the $B_c$-like molecular states, which may provide crucial information to disclose their inner structures.

Given the importance of the electromagnetic properties of the hadrons, various models and approaches have been adopted in the recent decades to discuss the magnetic moments of the hadronic states quantitatively \cite{Meng:2022ozq}. Borrowing the experience of studying the magnetic moments of the decuplet and octet baryons based on the constituent quark model \cite{Schlumpf:1993rm,Kumar:2005ei,Ramalho:2009gk}, the Lanzhou group already studied the electromagnetic properties of several heavy-flavor hadronic molecules within the constituent quark model in Refs. \cite{Zhou:2022gra,Wang:2022tib,Wang:2022ugk}, which can provide valuable information to reflect their inner structures.

In the present work, we first perform a quantitative calculation of the transition magnetic moments and the corresponding radiative decay behaviors of the $B_c$-like molecular states associated with their mass spectra \cite{Sun:2012sy}. Meanwhile, we also discuss the magnetic moments of the $B_c$-like molecular states. For achieving this goal, we adopt the constituent quark model, which is well established \cite{Liu:2003ab,Huang:2004tn,Zhu:2004xa,Haghpayma:2006hu,Wang:2016dzu,Deng:2021gnb,Gao:2021hmv,Zhou:2022gra,Wang:2022tib,Li:2021ryu,Schlumpf:1992vq,Schlumpf:1993rm,Cheng:1997kr,Ha:1998gf,Ramalho:2009gk,Girdhar:2015gsa,Menapara:2022ksj,Mutuk:2021epz,Menapara:2021vug,Menapara:2021dzi,Gandhi:2018lez,Dahiya:2018ahb,Kaur:2016kan,Thakkar:2016sog,Shah:2016vmd,Dhir:2013nka,Sharma:2012jqz,Majethiya:2011ry,Sharma:2010vv,Dhir:2009ax,Simonis:2018rld,
Ghalenovi:2014swa,Kumar:2005ei,Rahmani:2020pol,Hazra:2021lpa,Gandhi:2019bju,Majethiya:2009vx,Shah:2016nxi,Shah:2018bnr,Ghalenovi:2018fxh,Wang:2022ugk,Mohan:2022sxm,An:2022qpt,Kakadiya:2022pin,Wu:2022gie} and has been extensively used to discuss the magnetic moment properties of hadronic molecules in past decades \cite{Li:2021ryu,Zhou:2022gra,Wang:2022tib,Wang:2022ugk,Liu:2003ab,Wang:2016dzu,Deng:2021gnb,Gao:2021hmv}. In the concrete calculation, we discuss the role of the $S$-$D$ wave mixing effect to the electromagnetic properties of the isoscalar $D^{*}B^{*}$ and $D^{*}_{s}B^{*}_{s}$ molecules. By the present study, we may provide valuable information to disclose the properties of the $B_c$-like molecular states and give more abundant suggestions to search for the $B_c$-like molecular states in future experiments.

This paper is organized as follows. After the Introduction, we mainly study the transition magnetic moments and the corresponding radiative decay behaviors of the $B_c$-like molecular states in Sec. \ref{sec2}. In Sec. \ref{sec3}, we discuss the magnetic moments of the $B_c$-like molecular states, and answer whether or not the magnetic moment properties can be used to distinguish the hadrons with different configurations. Finally, this work ends with a short summary in Sec. \ref{sec4}.

\section{Transition magnetic moments and radiative decay behaviors}\label{sec2}

The study of the hadronic mass spectra not only provides crucial information to search for them experimentally, but also gives important input parameters for the study of their properties. In Ref. \cite{Sun:2012sy}, the Lanzhou group once predicted the mass spectra of the $S$-wave $B_c$-like molecules composed of the $D^{(*)}_{(s)}$ and $B^{(*)}_{(s)}$ mesons, where the $S$-$D$ wave mixing effect was considered. When the cutoff parameter in the form factor less than 2 GeV is a reasonable input value, the $DB$ state with $I(J^P)=0(0^+)$, the $DB^{*}$ state with $I(J^P)=0(1^+)$, the $D^{*}B$ state with $I(J^P)=0(1^+)$, the $D^{*}B^{*}$ states with $I(J^P)=0(0^+),\,0(1^+),\,0(2^+)$, and the $D^{*}_sB^{*}_s$ states with $I(J^P)=0(0^+),\,0(1^+),\,0(2^+)$ can be recommended as promising candidates of the $B_c$-like molecular states \cite{Sun:2012sy}.

As is well known, the electromagnetic properties of the hadrons mainly include the transition magnetic moments, the radiative decay behaviors, the magnetic moments, and so on. Compared with the magnetic moments of the $B_c$-like molecular states, it is very likely that the radiative decay behaviors of the $B_c$-like molecular states can be measured much more easily in future experiments. In the present work, we mainly study the transition magnetic moments and the corresponding radiative decay behaviors of the $B_c$-like molecular states. We hope that future experiments can bring us more surprises for the transition magnetic moments and the corresponding radiative decay behaviors of the $B_c$-like molecular states. In the concrete calculation, we adopt the constituent quark model and follow the same convention as previous work \cite{Li:2021ryu,Zhou:2022gra,Wang:2022tib,Wang:2022ugk} to calculate the transition magnetic moments and the corresponding radiative decay behaviors of the $B_c$-like molecular states.

First, we discuss the transition magnetic moments of the $B_c$-like molecular states within the constituent quark model, which can provide important input information for the study of the radiative decay behaviors of the $B_c$-like molecular states. Considering only the $S$-wave component, the transition magnetic moments of the $B_c$-like molecular states can be calculated by \cite{Wang:2022ugk}
\begin{eqnarray}
\mu_{H \to {H}^{\prime}}&=&\left\langle J_{H^{\prime}},J_{z} \left|\sum_{j}\hat{\mu}_{jz}^{\rm spin}e^{-i {\bf k}\cdot{\bf r}_j}\right| J_{H},J_{z} \right\rangle^{J_z={\rm Min}\{J_{H},\,J_{H^{\prime}}\}},\nonumber\\
\hat{\mu}_{jz}^{\rm spin}&=&\frac{e_j}{2m_j}\hat{\sigma}_{jz}.\label{EQMTBclike1}
\end{eqnarray}
Here, we use $e_j$, $m_j$, and $\hat{\sigma}_{jz}$ to represent the charge, the mass, and the $z$-component of the Pauli's spin operator of the $j$-th constituent of the hadron, respectively. In addition, we should mention that a lot of theoretical work use the maximum third component of the total angular momentum of the lowest state of the total angular momentum when presenting the transition magnetic moment of the hadrons \cite{Li:2021ryu,Zhou:2022gra,Wang:2022tib,Wang:2022ugk,Majethiya:2009vx,Majethiya:2011ry,Shah:2016nxi,Gandhi:2018lez,Simonis:2018rld,Ghalenovi:2018fxh,Gandhi:2019bju,Rahmani:2020pol,Hazra:2021lpa,Menapara:2021dzi,Menapara:2022ksj,Kakadiya:2022pin}. In the above expression, ${\bf k}$ is the momentum of the emitted photon with $k={(m_{H}^2-m_{H^{\prime}}^2)}/{2m_{H}}$. When the momentum of the emitted photon is close to zero and the spatial wave functions of the initial and final hadrons satisfy the normalization condition, the contribution of the factor $\left\langle R_{f} \left|e^{-i {\bf k}\cdot{\bf r}_j}\right| R_{i}\right\rangle$ can be ignored, and the above expression can be approximately written as
\begin{eqnarray}
\mu_{H \to {H}^{\prime}}=\left\langle J_{H^{\prime}},J_{z} \left|\sum_{j}\hat{\mu}_{jz}^{\rm spin}\right| J_{H},J_{z} \right\rangle^{J_z={\rm Min}\{J_{H},\,J_{H^{\prime}}\}}.\label{EQMTBclike2}
\end{eqnarray}
The previous work often use the above treatment approach to discuss the hadronic transition magnetic moments and radiative decay behaviors during the past few decades \cite{Majethiya:2009vx,Majethiya:2011ry,Shah:2016nxi,Gandhi:2018lez,Ghalenovi:2018fxh,Gandhi:2019bju,Rahmani:2020pol,Hazra:2021lpa,Menapara:2021dzi,Menapara:2022ksj,Li:2021ryu,Zhou:2022gra,Wang:2022tib,Kakadiya:2022pin}. However, the spatial wave functions of the initial and final states can influence the hadronic transition magnetic moments and radiative decay behaviors when the factor $\left\langle R_f \left|e^{-i {\bf k}\cdot{\bf r}_j}\right| R_i\right\rangle$ does not approach 1. Thus, we consider the contribution of the spatial wave functions of the initial and final states when discussing the transition magnetic moments and the radiative decay behaviors of the $B_c$-like molecular states in the following numerical analysis.

For the $B_c$-like molecular state, the color wave function is 1 due to the color confinement. Thus, the color wave functions do not affect the transition magnetic moments of the $B_c$-like molecular states \cite{Li:2021ryu,Zhou:2022gra,Wang:2022tib,Wang:2022ugk}. In the realistic calculation, it is necessary to discuss the spatial, flavor, and spin wave functions of the $B_c$-like molecular states and their constituent hadrons. In order to calculate the overlap of the spatial wave functions of the initial and final states quantitatively, we take the precise spatial wave functions for the $B_c$-like molecular states by solving the Schr\"odinger equation quantitatively \cite{Sun:2012sy}. In addition, we adopt the simple harmonic oscillator wave function $R_{n,l,m}(\beta,{\bf r})$ to describe the spatial wave functions of the mesons, which can be written as
\begin{eqnarray}\label{eq:sho}
R_{n,l,m}(\beta,{\bf r})&=&\sqrt{\frac{2n!}{\Gamma(n+l+\frac{3}{2})}}L_{n}^{l+\frac{1}{2}}(\beta^2r^2)\beta^{l+\frac{3}{2}}\nonumber\\
&&\times {\mathrm e}^{-\frac{\beta^2r^2}{2}}r^l Y_{l m}(\Omega).
\end{eqnarray}
Here, we use $L_{n}^{l+\frac{1}{2}}(x)$ and $Y_{l m}(\Omega)$ to denote the associated Laguerre polynomial and the spherical harmonic function, respectively. The radial, orbital, and magnetic quantum numbers of the mesons are marked as $n$, $l$, and $m$, respectively. In the concrete calculation, the oscillating parameters $\beta$ of the mesons can be estimated by fitting their mass spectra \cite{Workman:2022ynf}, and we can obtain $\beta_{D}=0.344~{\rm GeV}$, $\beta_{D^*}=0.291~{\rm GeV}$, $\beta_{B}=0.334~{\rm GeV}$, and $\beta_{B^*}=0.314~{\rm GeV}$. When calculating the factor $\left\langle R_f \left|e^{-i {\bf k}\cdot{\bf r}_j}\right| R_i\right\rangle$, we need to expand the spatial wave function of the emitted photon $e^{-i{\bf k}\cdot{\bf r}_j}$ by the spherical Bessel function $j_l(x)$ and the spherical harmonic function $Y_{l m}(\Omega)$, i.e., \cite{Khersonskii:1988krb}
\begin{eqnarray}
e^{-i{\bf k}\cdot{\bf r}_j}&=&\sum\limits_{l=0}^\infty\sum\limits_{m=-l}^l4\pi(-i)^lj_l(kr_j)Y_{lm}^*(\Omega_{\bf k})Y_{lm}(\Omega_{{\bf r}_j}).
\end{eqnarray}
With the above preparation, we can calculate the factor $\left\langle R_f \left|e^{-i {\bf k}\cdot{\bf r}_j}\right| R_i\right\rangle$.

Taking into account the coupling of the flavor wave functions of the constituent hadrons, the flavor wave functions $|I, I_3\rangle$ of the isoscalar $D^{(*)}_{(s)}B^{(*)}_{(s)}$ systems can be constructed as \cite{Sun:2012sy}
\begin{eqnarray}
D^{(*)}B^{(*)}&:& |0, 0\rangle=\frac{1}{\sqrt{2}}\left(D^{(*)0}B^{(*)+}+D^{(*)+}B^{(*)0}\right),\nonumber\\
D^{*}_{s}B^{*}_{s}&:& |0, 0\rangle=D^{*+}_{s}B^{*0}_{s},
\end{eqnarray}
where we use the notations $I$ and $I_3$ to denote the isospin and isospin's third component quantum numbers of the isoscalar $D^{(*)}_{(s)}B^{(*)}_{(s)}$ systems, respectively. The same method is applied to construct the spin wave functions $|S, S_3\rangle$ of the isoscalar $D^{(*)}_{(s)}B^{(*)}_{(s)}$ systems, which can be written as
\begin{eqnarray}
DB&:& |0, 0\rangle=|0, 0\rangle|0, 0\rangle,\nonumber\\
DB^*&:& |1, 1\rangle=|0, 0\rangle|1, 1\rangle,\nonumber\\
D^*B&:& |1, 1\rangle=|1, 1\rangle|0, 0\rangle,\nonumber\\
D^*_{(s)}B^*_{(s)}&:& |0, 0\rangle=\frac{1}{\sqrt{3}}\left(|1, 1\rangle|1, -1\rangle-|1, 0\rangle|1, 0\rangle\right.\nonumber\\
&&~~~~~~~~~~~\left.+|1, -1\rangle|1, 1\rangle\right),\nonumber\\
&& |1, 1\rangle=\frac{1}{\sqrt{2}}\left(|1, 1\rangle|1, 0\rangle-|1, 0\rangle|1, 1\rangle\right),\nonumber\\
&& |2, 2\rangle=|1, 1\rangle|1, 1\rangle.
\end{eqnarray}
Here, we take the notations $S$ and $S_3$ to represent the spin and spin's third component quantum numbers of the investigated hadrons, respectively.

Within the constituent quark model, the transition magnetic moments of the $B_c$-like molecular states can be related to the combination of the transition magnetic moments and the magnetic moments of their constituent hadrons. At present, the experimental information of the transition magnetic moments and the magnetic moments of the $D^{(*)}_{(s)}$ and $B^{(*)}_{(s)}$ mesons is lacking \cite{Workman:2022ynf}, but there exists a series of theoretical predictions. Thus, we can compare our obtained transition magnetic moments and magnetic moments of the $D^{(*)}_{(s)}$ and $B^{(*)}_{(s)}$ mesons with those from other theoretical work, which can test the reliability of our predicted results. Of course, we expect that the future experiments can focus on the transition magnetic moments and the magnetic moments of the $D^{(*)}_{(s)}$ and $B^{(*)}_{(s)}$ mesons, which can provide the important input information to improve our numerical analysis of the transition magnetic moments of the $S$-wave isoscalar $B_c$-like molecules.

In the present work, we estimate their transition magnetic moments and magnetic moments based on the constituent quark model. The flavor wave functions of the $D^{(*)}_{(s)}$ and $B^{(*)}_{(s)}$ mesons can be expressed as \cite{Workman:2022ynf,Simonis:2016pnh}
\begin{eqnarray}
(D^{(*)0},\,D^{(*)+},\,D_s^{(*)+})&=& (c\bar u,\,c\bar d,\,c\bar s),\nonumber\\
(B^{(*)+},\,B^{(*)0},\,B_s^{(*)0})&=& (u\bar b,\,d\bar b,\,s\bar b),
\end{eqnarray}
and their spin wave functions $|S,S_3\rangle$ can be constructed by the coupling of the spins of the constituent quarks, i.e.,
\begin{eqnarray}
D_{(s)}/B_{(s)}&:&\left|0,0\right\rangle=\dfrac{1}{\sqrt{2}}\left(\uparrow\downarrow-\downarrow\uparrow\right),\nonumber\\
D^{*}_{(s)}/B^{*}_{(s)}&:&\left|1,1\right\rangle=\uparrow\uparrow,\nonumber\\
&&\left|1,0\right\rangle=\dfrac{1}{\sqrt{2}}\left(\uparrow\downarrow+\downarrow\uparrow\right),\nonumber\\
&&\left|1,-1\right\rangle=\downarrow\downarrow,
\end{eqnarray}
where the notations $\uparrow$ and $\downarrow$ stand for the spin and the spin's third component quantum numbers of the quarks are $\left|{1}/{2},\,{1}/{2}\right\rangle$ and $\left|{1}/{2},\,-{1}/{2}\right\rangle$, respectively.

In order to represent the transition magnetic moments and the magnetic moments of the $D^{(*)}_{(s)}$ and $B^{(*)}_{(s)}$ mesons quantitatively, we adopt the constituent quark masses as $m_u=0.336~{\rm GeV}$, $m_d=0.336~{\rm GeV}$, $m_s=0.540~{\rm GeV}$, $m_c=1.660~{\rm GeV}$, and $m_b=4.730~{\rm GeV}$ in the following numerical analysis, which are often adopted to calculate the hadronic magnetic moments in the past decades \cite{Lichtenberg:1976fi,Li:2017cfz,Meng:2017dni,Li:2017pxa,Wang:2019mhm,Gao:2021hmv}. In Table \ref{TableTMMBc}, we present the numerical results of the transition magnetic moments of the $D^{(*)}$ and $B^{(*)}$ mesons. Here, we compare our predicted transition magnetic moments of the $D^{(*)}$ and $B^{(*)}$ mesons with those from other theoretical work, and our obtained transition magnetic moments of the $D^{(*)}$ and $B^{(*)}$ mesons are close to those from other theoretical work \cite{Zhou:2022gra,Simonis:2016pnh,Wang:2019mhm}. Thus, this fact reflects that our predicted transition magnetic moments of the $D^{(*)}$ and $B^{(*)}$ mesons are reliable within the constituent quark model.
\renewcommand\tabcolsep{0.45cm}
\renewcommand{\arraystretch}{1.50}
\begin{table}[!htbp]
  \caption{Our results of the transition magnetic moments of the $D^{(*)}$ and $B^{(*)}$ mesons and comparison with other results. Here, the transition magnetic moment of the hadrons is in unit of the nuclear magneton $\mu_N=e/2m_p$.}
  \label{TableTMMBc}
\begin{tabular}{c|c|c}
\toprule[1.0pt]\toprule[1.0pt]
Processes & Our works &Other works\\ \hline
$D^{*0} \to D^{0}$              & $2.134$ &  $2.233$ \cite{Zhou:2022gra},\,$2.250$ \cite{Simonis:2016pnh}\\
$D^{*+} \to D^{+}$              & $-0.515$ &  $-0.559$ \cite{Zhou:2022gra},\,$-0.540$ \cite{Simonis:2016pnh}\\
$B^{*0} \to B^{0}$              & $-0.990$ &  $-0.990$ \cite{Simonis:2016pnh},\,$-1.00$ \cite{Wang:2019mhm}\\
$B^{*+} \to B^{+}$              & $1.783$ &  $1.800$ \cite{Simonis:2016pnh},\,$1.730$ \cite{Wang:2019mhm}\\
\bottomrule[1.0pt]
\bottomrule[1.0pt]
\end{tabular}
\end{table}

For the $S$-wave $D^{(*)}_{(s)}$ and $B^{(*)}_{(s)}$ mesons, there only exist the spin magnetic moments. According to Refs. \cite{Liu:2003ab,Huang:2004tn,Zhu:2004xa,Haghpayma:2006hu,Wang:2016dzu,Deng:2021gnb,Gao:2021hmv,Zhou:2022gra,Wang:2022tib,Li:2021ryu,Schlumpf:1992vq,Schlumpf:1993rm,Cheng:1997kr,Ha:1998gf,Ramalho:2009gk,Girdhar:2015gsa,
Menapara:2022ksj,Mutuk:2021epz,Menapara:2021vug,Menapara:2021dzi,Gandhi:2018lez,Dahiya:2018ahb,Kaur:2016kan,Thakkar:2016sog,Shah:2016vmd,Dhir:2013nka,Sharma:2012jqz,Majethiya:2011ry,Sharma:2010vv,Dhir:2009ax,Simonis:2018rld,
Ghalenovi:2014swa,Kumar:2005ei,Rahmani:2020pol,Hazra:2021lpa,Gandhi:2019bju,Majethiya:2009vx,Shah:2016nxi,Shah:2018bnr,Ghalenovi:2018fxh,Wang:2022ugk,Mohan:2022sxm,An:2022qpt,Kakadiya:2022pin,Wu:2022gie}, the spin magnetic moments of the hadrons can be calculated by the following expectation values:
\begin{eqnarray}
\mu_{{H}}=\left\langle J_{H},J_{H} \left|\sum_{j}\hat{\mu}_{jz}^{\rm spin}\right| J_{H},J_{H} \right\rangle.\label{EQMBclike}
\end{eqnarray}
Here, we need to specify that a lot of theoretical work use the maximum third component of the total angular momentum of the hadron when presenting the magnetic moment of the hadron \cite{Liu:2003ab,Huang:2004tn,Zhu:2004xa,Haghpayma:2006hu,Wang:2016dzu,Deng:2021gnb,Gao:2021hmv,Zhou:2022gra,Wang:2022tib,Li:2021ryu,Schlumpf:1992vq,Schlumpf:1993rm,Cheng:1997kr,Ha:1998gf,Ramalho:2009gk,Girdhar:2015gsa,
Menapara:2022ksj,Mutuk:2021epz,Menapara:2021vug,Menapara:2021dzi,Gandhi:2018lez,Dahiya:2018ahb,Kaur:2016kan,Thakkar:2016sog,Shah:2016vmd,Dhir:2013nka,Sharma:2012jqz,Majethiya:2011ry,Sharma:2010vv,Dhir:2009ax,Simonis:2018rld,
Ghalenovi:2014swa,Kumar:2005ei,Rahmani:2020pol,Hazra:2021lpa,Gandhi:2019bju,Majethiya:2009vx,Shah:2016nxi,Shah:2018bnr,Ghalenovi:2018fxh,Wang:2022ugk,Mohan:2022sxm,An:2022qpt,Kakadiya:2022pin,Wu:2022gie}. In the realistic calculation, the spatial wave functions of the $S$-wave $D^{(*)}_{(s)}$ and $B^{(*)}_{(s)}$ mesons satisfy the normalization condition. Thus, the spatial wave functions of the $S$-wave $D^{(*)}_{(s)}$ and $B^{(*)}_{(s)}$ mesons do not affect their magnetic moments \cite{Li:2021ryu,Zhou:2022gra,Wang:2022tib,Wang:2022ugk}.

In Table \ref{TableMT}, we present the results of the magnetic moments of the $D^{(*)}_{(s)}$ and $B^{(*)}_{(s)}$ mesons. For the $S$-wave charmed meson $D$ and bottom meson $B$, their magnetic moments are zero, which is due to the fact that their spin quantum numbers are zero. In addition, we also present the magnetic moments of the $D^{(*)}_{(s)}$ and $B^{(*)}_{(s)}$ mesons from other theoretical work in Table \ref{TableMT}. By comparing these numerical results, we find that our obtained magnetic moments of the $D^{(*)}_{(s)}$ and $B^{(*)}_{(s)}$ mesons are close to those from other theoretical predictions \cite{Zhou:2022gra,Simonis:2018rld,Simonis:2016pnh,Wang:2019mhm,Luan:2015goa}, which may reflect our adopted constituent quark masses are reliable.

\renewcommand\tabcolsep{0.25cm}
\renewcommand{\arraystretch}{1.50}
\begin{table}[!htbp]
  \caption{Our results of the magnetic moments of the $D^{(*)}_{(s)}$ and $B^{(*)}_{(s)}$ mesons and comparison with other results. Here, we define $\mu_q=e_q/2m_q$, and the magnetic moment of the hadron is in unit of the nuclear magneton $\mu_N$.}
  \label{TableMT}
\begin{tabular}{c|cc|c}
\toprule[1.0pt]\toprule[1.0pt]
\multirow{2}{*}{Mesons} & \multicolumn{2}{c|}{Our work} &\multirow{2}{*}{Other work}\\
                        & Expressions      & Values        &  \\ \hline
$D^{*0}$                &$\mu_{c}+\mu_{\bar u}$ & $-1.485$ &  $-1.489$ \cite{Zhou:2022gra},\,$-1.470$ \cite{Simonis:2016pnh}\\
$D^{*+}$                &$\mu_{c}+\mu_{\bar d}$ & $1.308$ &  $1.303$ \cite{Zhou:2022gra},\,$1.320$ \cite{Simonis:2016pnh}\\
$D_s^{*+}$                &$\mu_{c}+\mu_{\bar s}$ & $0.956$ &  $0.930$ \cite{Simonis:2018rld},\,$0.980$ \cite{Luan:2015goa}\\
$B^{*0}$                &$\mu_{d}+\mu_{\bar b}$ & $-0.865$ &  $-0.870$ \cite{Simonis:2016pnh},\,$-0.870$ \cite{Wang:2019mhm}\\
$B^{*+}$                &$\mu_{u}+\mu_{\bar b}$ & $1.928$ &  $1.920$ \cite{Simonis:2016pnh},\,$1.900$ \cite{Wang:2019mhm}\\
$B_s^{*0}$                &$\mu_{s}+\mu_{\bar b}$ & $-0.513$ &  $-0.513$ \cite{Simonis:2018rld},\,$-0.550$ \cite{Simonis:2016pnh}\\
\bottomrule[1.0pt]
\bottomrule[1.0pt]
\end{tabular}
\end{table}

Because the hadronic molecular state is a loosely bound state with the reasonable binding energy at most tens of MeV \cite{Chen:2016qju}, the masses of the isoscalar $D^{(*)}B^{(*)}$ molecular states should satisfy the relation $m_{DB}<m_{DB^*}<m_{D^*B}<m_{D^*B^*}$. Furthermore, the binding energies of the isoscalar $D^*B^*$ states are $E_{D^*B^*(2^+)}<E_{D^*B^*(1^+)}<E_{D^*B^*(0^+)}$ when taking the same cutoff values \cite{Sun:2012sy}, which shows that their masses satisfy the relation $m_{D^*B^*(2^+)}<m_{D^*B^*(1^+)}<m_{D^*B^*(0^+)}$. Thus, the masses of the isoscalar $D^{(*)}B^{(*)}$ molecular states are $m_{DB(0^+)}<m_{DB^*(1^+)}<m_{D^*B(1^+)}<m_{D^*B^*(2^+)}<m_{D^*B^*(1^+)}<m_{D^*B^*(0^+)}$. Similar to the above analysis, the binding energies of the isoscalar $D_s^{*}B_s^{*}$ states satisfy $E_{D_s^{*}B_s^{*}(0^+)}<E_{D_s^{*}B_s^{*}(1^+)}<E_{D_s^{*}B_s^{*}(2^+)}$ if adopting the same cutoff parameters \cite{Sun:2012sy}. Thus, the masses of the  isoscalar $D_s^{*}B_s^{*}$ molecular states are $m_{D_s^*B_s^*(0^+)}<m_{D_s^*B_s^*(1^+)}<m_{D_s^*B_s^*(2^+)}$.

Assuming the mass relations of the isoscalar $D^{(*)}_{(s)}B^{(*)}_{(s)}$ molecular states obtained above, we can further discuss their transition magnetic moments. When considering the contribution of the spatial wave functions of the initial and final states, we take the precise spatial wave functions for the $B_c$-like molecular states in this work, which can be obtained by the study of their mass spectra \cite{Sun:2012sy}. However, the precise spatial wave functions of the $B_c$-like molecular states depend on their binding energies, and the experimental date of the binding energies of the $B_c$-like molecular states is lacking up to now \cite{Workman:2022ynf}. For simplicity, we take the same binding energies for the initial and final $B_c$-like molecular states and use $-2$, $-7$, and $-12~{\rm MeV}$ to discuss their transition magnetic moments in this work. In Table \ref{TableTMMBclike}, the numerical results of the transition magnetic moments of the $S$-wave isoscalar $B_c$-like molecular states are collected.

\renewcommand\tabcolsep{0.28cm}
\renewcommand{\arraystretch}{1.50}
\begin{table}[!htbp]
  \caption{The transition magnetic moments of the $S$-wave isoscalar $B_c$-like molecular states when only considering the $S$-wave component. Here, the transition magnetic moment of the hadrons is in unit of the nuclear magneton $\mu_N$.}
  \label{TableTMMBclike}
\begin{tabular}{c|ccc}
\toprule[1.0pt]\toprule[1.0pt]
\multirow{2}{*}{Processes}   &\multicolumn{3}{c}{Transition magnetic moments}\\
                             &$-2~{\rm MeV}$&$-7~{\rm MeV}$&$-12~{\rm MeV}$\\\hline
{$DB^*(1^+) \to DB(0^+)$}&0.395&0.396&0.396\\
{$D^*B(1^+) \to DB(0^+)$}&0.684&0.758&0.773\\
{$D^*B^*(2^+) \to DB^*(1^+)$}&0.486&0.537&0.547\\
{$D^*B^*(1^+) \to DB^*(1^+)$}&$-0.484$&$-0.536$&$-0.546$\\
{$D^*B^*(0^+) \to DB^*(1^+)$}&$-0.394$&$-0.435$&$-0.442$\\
{$D^*B^*(2^+) \to D^*B(1^+)$}&0.279&0.280&0.280\\
{$D^*B^*(1^+) \to D^*B(1^+)$}&0.279&0.279&0.279\\
{$D^*B^*(0^+) \to D^*B(1^+)$}&$-0.227$&$-0.226$&$-0.225$\\
{$D^*B^*(1^+) \to D^*B^*(2^+)$}&$-0.310$&$-0.310$&$-0.310$\\
{$D^*B^*(0^+) \to D^*B^*(1^+)$}&$-0.506$&$-0.506$&$-0.506$\\
{$D_s^*B_s^*(1^+) \to D_s^*B_s^*(0^+)$}&$1.199$&$1.199$&$1.199$\\
{$D_s^*B_s^*(2^+) \to D_s^*B_s^*(1^+)$}&0.735&0.735&0.735\\
\bottomrule[1.0pt]\bottomrule[1.0pt]
\end{tabular}
\end{table}

As presented in Table \ref{TableTMMBclike}, the largest transition magnetic moment is $1.199\mu_N$ and corresponds to the $D_s^*B_s^*(1^+) \to D_s^*B_s^*(0^+) \gamma$ process, while the $D^*B^*(0^+) \to D^*B^*(1^+) \gamma$ process has the smallest transition magnetic moment and is $-0.506\mu_N$. In addition, the transition magnetic moments of the $D^*B^*(1^+) \to D^*B^*(2^+)$, $D^*B^*(0^+) \to D^*B^*(1^+)$, $D_s^*B_s^*(1^+) \to D_s^*B_s^*(0^+)$, and $D_s^*B_s^*(2^+) \to D_s^*B_s^*(1^+)$ processes are not affected by the binding energies for the initial and final $B_c$-like molecular states, since the factor $\left\langle R_f \left|e^{-i {\bf k}\cdot{\bf r}_j}\right| R_i\right\rangle$ is extremely close to 1 for these radiative decay processes when taking the same binding energies for the initial and final $B_c$-like molecular states.

Then, we further discuss the radiative decay behaviors of the $S$-wave isoscalar $B_c$-like molecular states. According to Refs. \cite{Dey:1994qi,Simonis:2018rld,Gandhi:2019bju,Hazra:2021lpa,Li:2021ryu,Zhou:2022gra,Wang:2022tib,Rahmani:2020pol,Menapara:2022ksj,Menapara:2021dzi,Gandhi:2018lez,Majethiya:2011ry,Majethiya:2009vx,Shah:2016nxi,Ghalenovi:2018fxh,Wang:2022ugk,Mohan:2022sxm,An:2022qpt,Kakadiya:2022pin}, the widths for the radiative decay processes $B_c \to B_c^{\prime}\gamma$ can be expressed in terms of the transition magnetic moments, which can be given by \cite{Wang:2022ugk}
\begin{eqnarray}
 \Gamma_{{B_c} \to B_c^{\prime}\gamma}=\frac{k^{3}}{m_{p}^{2}} \frac{\alpha_{\rm {EM}}}{2J_{{B_c}}+1}\frac{\sum\limits_{J_{B_c^{\prime}z},J_{{B_c}z}}\left(\begin{array}{ccc} J_{B_c^{\prime}}&1&J_{{B_c}}\\-J_{B_c^{\prime}z}&0&J_{{B_c}z}\end{array}\right)^2}{\left(\begin{array}{ccc} J_{B_c^{\prime}}&1&J_{{B_c}}\\-J_{z}&0&J_{z}\end{array}\right)^2}\frac{\left|\mu_{{B_c} \to B_c^{\prime}}\right|^2}{\mu_N^2}.\nonumber\\
\end{eqnarray}
Here, $k$ represents the momentum of the emitted photon with $k={(m_{B_c}^2-m_{B_c^{\prime}}^2)}/{2m_{B_c}}$, $m_p$ is the mass of the proton with $m_p=0.938~{\rm{GeV}}$ \cite{Workman:2022ynf}, the electromagnetic fine structure constant $\alpha_{\rm {EM}}$ is taken as $\alpha_{\rm {EM}} \approx {1}/{137}$, the notation $\left(\begin{array}{ccc} a&b&c\\d&e&f\end{array}\right)$ is the 3-$j$ coefficient, and $J_z={\rm Min}\{J_{B_c},\,J_{B_c^{\prime}}\}$. As mentioned above, we also assume the same binding energies for the initial and final $B_c$-like molecular states and take $-2$, $-7$, and $-12~{\rm MeV}$ to discuss these radiative decay widths in this work. In Table \ref{TableRDBBclike}, the radiative decay widths of the $S$-wave isoscalar $B_c$-like molecular states are collected.
\renewcommand\tabcolsep{0.28cm}
\renewcommand{\arraystretch}{1.50}
\begin{table}[!htbp]
  \caption{The radiative decay widths of the $S$-wave isoscalar $B_c$-like molecular states when only considering the $S$-wave component. Here, the radiative decay width of the hadrons is in unit of ${\rm keV}$.}
  \label{TableRDBBclike}
\begin{tabular}{c|ccc}
\toprule[1.0pt]\toprule[1.0pt]
\multirow{2}{*}{Processes}   &\multicolumn{3}{c}{Radiative decay widths}\\
                             &$-2~{\rm MeV}$&$-7~{\rm MeV}$&$-12~{\rm MeV}$\\\hline
{$DB^*(1^+) \to DB(0^+) \gamma$}    &0.039         &0.040         &0.040\\
{$D^*B(1^+) \to DB(0^+) \gamma$}    &3.542         &4.358         &4.527\\
{$D^*B^*(2^+) \to DB^*(1^+) \gamma$}&3.578         &4.366         &4.529\\
{$D^*B^*(1^+) \to DB^*(1^+) \gamma$}&3.547         &4.359         &4.521\\
{$D^*B^*(0^+) \to DB^*(1^+) \gamma$}&3.536         &4.230         &4.434\\
{$D^*B^*(2^+) \to D^*B(1^+) \gamma$}&0.039          &0.040          &0.040 \\
{$D^*B^*(1^+) \to D^*B(1^+) \gamma$}&0.039          &0.040          &0.039\\
{$D^*B^*(0^+) \to D^*B(1^+) \gamma$}&0.039          &0.039          &0.039\\
\bottomrule[1.0pt]\bottomrule[1.0pt]
\end{tabular}
\end{table}

According to Table \ref{TableRDBBclike}, the radiative decay widths of the $B_c$-like molecular states strongly depend on the transition magnetic moments and the phase spaces. For example, the radiative decay width of the $D^*B(1^+) \to DB(0^+) \gamma$ process is larger than that of the $DB^*(1^+) \to DB(0^+) \gamma$ process, since the $D^*B(1^+) \to DB(0^+) \gamma$ process has more transition magnetic moment and phase space compared with the $DB^*(1^+) \to DB(0^+) \gamma$ process. The same reason also leads to the radiative decay behaviors between the $D^*B^*(2^+/1^+/0^+) \to DB^*(1^+) \gamma$ and $D^*B^*(2^+/1^+/0^+) \to D^*B(1^+) \gamma$ processes. In addition, several radiative decay processes have significant widths, which may provide the crucial information to an experimental search for the $B_c$-like molecular states.

When adopting the same binding energies for the initial and final $B_c$-like molecular states, the radiative decay widths are zero for the $D^*B^*(1^+) \to D^*B^*(2^+) \gamma$, $D^*B^*(0^+) \to D^*B^*(1^+) \gamma$, $D_s^*B_s^*(1^+) \to D_s^*B_s^*(0^+) \gamma$, and $D_s^*B_s^*(2^+) \to D_s^*B_s^*(1^+) \gamma$ processes, which is because the phase spaces are zero for these radiative decay processes. However, the initial and final $B_c$-like molecules may have different binding energies for these radiative decay processes. In the following analysis, we take different binding energies for the initial and final $B_c$-like molecular states to discuss the radiative decay widths of the $D^*B^*(1^+) \to D^*B^*(2^+) \gamma$, $D^*B^*(0^+) \to D^*B^*(1^+) \gamma$, $D_s^*B_s^*(1^+) \to D_s^*B_s^*(0^+) \gamma$, and $D_s^*B_s^*(2^+) \to D_s^*B_s^*(1^+) \gamma$ processes. By scanning the binding energies of the initial and final $B_c$-like molecular states in the range $-12$ to $-2~{\rm MeV}$, we can further estimate the following relations
\begin{eqnarray}
&&\Gamma_{D^*B^*(1^+) \to D^*B^*(2^+) \gamma} < 0.001~{\rm keV},\nonumber\\
&&\Gamma_{D^*B^*(0^+) \to D^*B^*(1^+) \gamma} < 0.002~{\rm keV},\nonumber\\
&&\Gamma_{D_s^*B_s^*(1^+) \to D_s^*B_s^*(0^+) \gamma} < 0.004~{\rm keV},\nonumber\\
&&\Gamma_{D_s^*B_s^*(2^+) \to D_s^*B_s^*(1^+) \gamma} < 0.003~{\rm keV}.
\end{eqnarray}
Obviously, the widths of these radiative decay processes are strongly suppressed, which is because the masses of the initial and final $B_c$-like molecular states are extremely close to each other for these radiative decay processes.

In the following, we discuss the role of the $S$-$D$ wave mixing effect to the transition magnetic moments of the isoscalar $D^{*}B^{*}$ and $D^{*}_{s}B^{*}_{s}$ molecules, which is similar to the study of the mass spectra of the $B_c$-like molecular states in Ref. \cite{Sun:2012sy}. Before calculating the transition magnetic moments of the $B_c$-like molecular states after including the $S$-$D$ wave mixing effect, we simply review their mass spectra in the following \cite{Sun:2012sy}:
\begin{enumerate}
  \item For the $DB$ state with $I(J^P)=0(0^+)$, there only exists the $|{}^1S_0\rangle$ channel.
  \item For the $DB^{*}$ and $D^{*}B$ states with $I(J^P)=0(1^+)$, the probabilities of the $D$-wave channels are zero, which is due to the absence of the contribution of the tensor forces from the $S$-$D$ wave mixing effect for the $DB^{*}$ and $D^{*}B$ interactions. Thus, the $D$-wave channels do not affect their magnetic moment properties.
  \item For the $D^{*}B^{*}$ and $D^{*}_sB^{*}_s$ states with $I(J^P)=0(0^+),\,0(1^+),\,0(2^+)$, the contribution from the $S$-$D$ wave mixing effect can affect their mass spectra. Similarly, we conjecture that the $D$-wave channels can influence their magnetic moment properties.
\end{enumerate}
In this work, we take into account the following $S$-wave and $D$-wave channels for the isoscalar $D^{*}B^{*}$ and $D^{*}_{s}B^{*}_{s}$ states with $J^P=0^+,\,1^+,\,2^+$ \cite{Sun:2012sy}, i.e.,
\begin{eqnarray}
&&D^{*}B^{*}/D^{*}_{s}B^{*}_{s}(0^+):~~|^{1} S_{0}\rangle,\,|^{5} D_{0}\rangle,\nonumber\\
&&D^{*}B^{*}/D^{*}_{s}B^{*}_{s}(1^+):~~|^{3} S_{1}\rangle,\,|^{3} D_{1}\rangle,\,|^{5} D_{1}\rangle,\nonumber\\
&&D^{*}B^{*}/D^{*}_{s}B^{*}_{s}(2^+):~~|^{5} S_{2}\rangle,\,|^{1} D_{2}\rangle,\,|^{3} D_{2}\rangle,\,|^{5} D_{2}\rangle.
\end{eqnarray}
Here, we take the notation $|^{2S+1} L_{J}\rangle$ to mark the quantum numbers of the corresponding channel, while $S$, $L$, and $J$ denote the spin, orbit angular momentum, and total angular momentum quantum numbers, respectively.

When discussing the transition magnetic moments of the isoscalar $D^{*}B^{*}$ and $D^{*}_{s}B^{*}_{s}$ molecules after considering the $S$-$D$ wave mixing effect, it is necessary to deduce the transition magnetic moments of the corresponding $S$-wave and $D$-wave channels. For the $D_{(s)}^*B_{(s)}^*(2^+) \to D_{(s)}^*B_{(s)}^*(0^+)$, $D_{(s)}^*B_{(s)}^*(1^+) \to D_{(s)}^*B_{(s)}^*(0^+)$, and $D_{(s)}^*B_{(s)}^*(2^+) \to D_{(s)}^*B_{(s)}^*(1^+)$ processes, we need to mention that the momenta of the emitted photon are zero when adopting the same binding energies for the initial and final $B_c$-like molecular states, and their transition magnetic moments can be calculated by the following expectation values \cite{Wang:2022ugk}:
\begin{eqnarray}
\mu_{H \to {H}^{\prime}}&=&\left\langle J_{H^{\prime}},J_{z} \left|\sum_{j}\hat{\mu}_{jz}^{\rm spin}+\hat{\mu}_{z}^{\rm orbital} \right| J_{H},J_{z} \right\rangle^{J_z={\rm Min}\{J_{H},\,J_{H^{\prime}}\}},\nonumber\\
\hat{\mu}_{z}^{\rm orbital}&=&\left(\frac{m_{\alpha}}{m_{\alpha}+m_{\beta}}\frac{e_{\beta}}{2m_{\beta}}+\frac{m_{\beta}}{m_{\alpha}+m_{\beta}}\frac{e_{\alpha}}{2m_{\alpha}}\right)\hat{L}_z.
\label{MBclikeSD}
\end{eqnarray}
For the $D^{*}B^{*}$ and $D^{*}_{s}B^{*}_{s}$ systems, the notations $\alpha$ and $\beta$ denote the $D^{*}_{(s)}$ and $B^{*}_{(s)}$ mesons, and $\hat{L}_z$ is the $z$-component of the orbital angular momenta operator between the $D^{*}_{(s)}$ and $B^{*}_{(s)}$ mesons. In the present work, we take the masses of the $D^{(*)}_{(s)}$ and $B^{(*)}_{(s)}$ mesons from the Particle Data Group \cite{Workman:2022ynf}.

For obtaining the transition magnetic moments of the $D$-wave channels, we follow the standard strategy in Refs. \cite{Li:2021ryu,Zhou:2022gra,Wang:2022tib,Wang:2022ugk}. In the realistic calculation, we first expand their spin-orbital wave functions $|{ }^{2 S+1} L_{J}\rangle$ by the orbital wave function $Y_{L, m_{L}}$ and the spin wave function $\chi_{S,\,m_{S}}$, i.e.,
\begin{eqnarray}
\left|{ }^{2 S+1} L_{J}\right\rangle=\sum_{m_{L},m_{S}} C_{L m_{L},S m_{S}}^{J M} Y_{L, m_{L}}\chi_{S,\,m_{S}}.
\end{eqnarray}
Here, $C_{L m_{L},S m_{S}}^{J M}$ is the Clebsch-Gordan coefficient. Then, the transition magnetic moments of these discussed $D$-wave channels can be deduced by calculating the expectation values of the spin and orbital magnetic moment operators.

Based on the obtained transition magnetic moments of the $S$-wave and $D$-wave channels, the transition magnetic moments of the isoscalar $D^{*}B^{*}$ and $D^{*}_{s}B^{*}_{s}$ molecules can be calculated by summing the contribution of the transition magnetic moments of the corresponding mixing channels. Here, we need to mention that the transition magnetic moments of the hadronic molecules not only depend on the transition magnetic moments of the relevant $S$-wave and $D$-wave channels, but also rely on the spatial wave functions of the mixing channels when considering the contribution of the $S$-$D$ wave mixing effect. However, the spatial wave functions of the isoscalar $D^{*}B^{*}$ and $D^{*}_{s}B^{*}_{s}$ molecules are related to their binding energies. Here, we also take three typical binding energies $-2$, $-7$, and $-12~{\rm MeV}$ for the isoscalar $D^{*}B^{*}$ and $D^{*}_{s}B^{*}_{s}$ molecules to discuss these transition magnetic moments when considering the contribution of the $S$-$D$ wave mixing effect. In Table \ref{TableTMBcSDResults}, we present the transition magnetic moments of the isoscalar $D^{*}B^{*}$ and $D^{*}_{s}B^{*}_{s}$ molecular states after including the $S$-$D$ wave mixing effect.

\renewcommand\tabcolsep{0.30cm}
\renewcommand{\arraystretch}{1.50}
\begin{table}[!htbp]
  \caption{The transition magnetic moments of the isoscalar $D^{*}B^{*}$ and $D^{*}_{s}B^{*}_{s}$ molecules after including the $S$-$D$ wave mixing effect. Here, the transition magnetic moment of the hadrons is in unit of the nuclear magneton $\mu_N$.}
  \label{TableTMBcSDResults}
\begin{tabular}{c|ccc}
\toprule[1.0pt]\toprule[1.0pt]
\multirow{2}{*}{Processes}&\multicolumn{3}{c}{Transition magnetic moments}\\
                         &$-2~{\rm MeV}$&$-7~{\rm MeV}$&$-12~{\rm MeV}$\\\hline
$D^*B^*(1^+) \to D^*B^*(2^+)$                &$-0.307$   &$-0.306$  &$-0.306$\\
$D^*B^*(0^+) \to D^*B^*(1^+)$                &$-0.481$   &$-0.467$ &$-0.460$\\
$D_s^*B_s^*(1^+) \to D_s^*B_s^*(0^+)$        &1.185      &1.174    &1.169\\
$D_s^*B_s^*(2^+) \to D_s^*B_s^*(1^+)$        &0.731      &0.728    &0.726\\
\bottomrule[1.0pt]\bottomrule[1.0pt]
\end{tabular}
\end{table}

As shown in Table \ref{TableTMBcSDResults}, the $D$-wave channels with a small contribution \cite{Sun:2012sy} play a minor role to decorate the transition magnetic moments of the isoscalar $D^{*}B^{*}$ and $D^{*}_{s}B^{*}_{s}$ molecules. When considering the contribution of the $D$-wave channels, the change of their transition magnetic moments is less than $0.05\mu_N$, and the most obvious change is the $D^*B^*(0^+) \to D^*B^*(1^+) \gamma$ process. Additionally, the radiative decay widths of the isoscalar $D^{*}B^{*}$ and $D^{*}_{s}B^{*}_{s}$ molecules depend on their transition magnetic moments. Thus, the $S$-$D$ wave mixing effect also plays a minor role to change the radiative decay widths of the isoscalar $D^{*}B^{*}$ and $D^{*}_{s}B^{*}_{s}$ molecular states.

As the important input parameters, the information of the constituent quark masses is crucial when discussing the transition magnetic moments and the radiative decay behaviors of the $B_c$-like molecular states within the constituent quark model. However, the constituent quark masses cannot be accurately determined due to the lack of relevant experimental data. In order to intuitively clarify the reliability of our adopted constituent quark masses, we have compared our obtained transition magnetic moments and magnetic moments of the $D^{(*)}_{(s)}$ and $B^{(*)}_{(s)}$ mesons with those from other theoretical work, and we find that our obtained numerical results are comparable with those from other theoretical work, which may reflect that our adopted constituent quark masses are relatively reliable.

In the following, we briefly discuss the theoretical errors for the transition magnetic moments and the radiative decay widths of the $S$-wave isoscalar $B_c$-like molecular states. By comparing our adopted constituent quark masses with those of Ref. \cite{Majethiya:2009vx}, we find that the differences in the constituent quark masses are less than 10\%. To estimate the theoretical errors for the transition magnetic moments and the radiative decay widths of the $S$-wave isoscalar $B_c$-like molecular states, we take 10\% uncertainties for the constituent quark masses. Since the transition magnetic moments of the hadrons are inversely proportional to the constituent quark masses, we can obtain
\begin{eqnarray}
\frac{1}{m_q+\delta m_q}=\frac{1}{m_q\left(1+\frac{\delta m_q}{m_q}\right)}\approx \frac{1}{m_q}\left(1-\frac{\delta m_q}{m_q}\right)=\frac{1}{m_q}\left(1-0.1\right).\nonumber\\
\end{eqnarray}
Here, we take the uncertainties of the constituent quark masses $\delta m_q=0.1m_q$, reflecting that $\delta m_q/m_q$ is the small amount compared to 1. Thus, the theoretical errors for the transition magnetic moments of the $S$-wave isoscalar $B_c$-like molecular states are about $10\%$, assuming 10\% uncertainties for the constituent quark masses. Since the radiative decay widths of the hadrons are proportional to the squares of the transition magnetic moments of the hadrons, we have
\begin{eqnarray}
\left(\mu+\delta \mu\right)^2=\mu^2\left(1+\frac{\delta \mu}{\mu}\right)^2\approx\mu^2\left(1+0.2\right),
\end{eqnarray}
where we take the uncertainties of the transition magnetic moments $\delta \mu=0.1\mu$. Thus, the theoretical errors for the radiative decay widths of the $S$-wave isoscalar $B_c$-like molecular states are around $20\%$, assuming 10\% uncertainties for the constituent quark masses. {Undoubtedly, our anticipation lies in the forthcoming experiments and lattice QCD simulations, which will predominantly concentrate on investigating the transition magnetic moments and radiative decay widths of hadronic molecules. With the data acquired from these investigations, we can refine the constituent quark masses within the constituent quark model. This refined information will serve as crucial input for analyzing the transition magnetic moments and radiative decay characteristics of isoscalar $B_c$-like molecular states.}

{In addition to constituent quark masses, the determination of transition magnetic moments and radiative decay widths of hadrons in the constituent quark model relies on various factors, including spatial wave functions and the masses of the initial and final states. Consequently, uncertainties associated with these factors introduce theoretical errors when calculating the transition magnetic moments and radiative decay widths of isoscalar $B_c$-like molecular states. Notably, the spatial wave functions and masses of the isoscalar $B_c$-like molecules are interconnected with their binding energies.
Presently, the experimental detection of isoscalar $B_c$-like molecular states is lacking. In our current study, we consider three representative binding energies to investigate the transition magnetic moments and radiative decay widths. The numerical results for these quantities, as presented in Tables \ref{TableTMMBclike} and \ref{TableRDBBclike}, indicate that the uncertainties arising from the spatial wave functions and the masses of the initial and final states contribute to theoretical errors of less than 15\% for the transition magnetic moments and less than 30\% for the radiative decay widths of the isoscalar $B_c$-like molecular states.
Therefore, we express strong anticipation for future experimental efforts aimed at measuring the binding energies of isoscalar $B_c$-like molecules. Such measurements would significantly enhance our understanding of the transition magnetic moments and radiative decay behaviors of these intriguing molecular states.}

\section{Magnetic moment properties}\label{sec3}

In this section, we discuss the magnetic moments of the $B_c$-like molecular states, and answer whether or not the magnetic moment properties can be used to distinguish the hadrons with different configurations.

\subsection{Magnetic moments of the $B_c$-like molecules}

In order to disclose the properties of the $B_c$-like molecules, in this subsection we study the magnetic moment properties of the $B_c$-like molecular states associated with their mass spectra. In the concrete calculation, we adopt the constituent quark model and follow the same convention as the previous work \cite{Li:2021ryu,Zhou:2022gra,Wang:2022tib,Wang:2022ugk} to calculate the magnetic moments of the $B_c$-like molecular states. In our numerical analysis, we discuss the magnetic moments of the $B_c$-like molecular states by performing the single channel and $S$-$D$ wave mixing analysis, respectively.

First, we discuss the magnetic moments of the $B_c$-like molecular states when only considering the $S$-wave component, and there only exist the spin magnetic moments. In the realistic calculation, the spatial wave function satisfies the normalization condition when only considering the contribution of the $S$-wave component. Thus, the spatial wave functions do not affect the magnetic moments of the $B_c$-like molecular states when only focusing on the contribution of the $S$-wave component \cite{Li:2021ryu,Zhou:2022gra,Wang:2022tib,Wang:2022ugk}.

By calculating Eq. (\ref{EQMBclike}), we can obtain the magnetic moments of the $S$-wave isoscalar $B_c$-like molecular states, which can be written as
\begin{eqnarray}
\mu_{DB(0^+)}&=&0,\nonumber\\
\mu_{DB^*(1^+)}&=&\frac{1}{2}\left(\mu_{B^{*+}}+\mu_{B^{*0}}\right),\nonumber\\
\mu_{D^*B(1^+)}&=&\frac{1}{2}\left(\mu_{D^{*+}}+\mu_{D^{*0}}\right),\nonumber\\
\mu_{D^*B^*(0^+)}&=&0,\nonumber\\
\mu_{D^*B^*(1^+)}&=&\frac{1}{4}\left(\mu_{D^{*+}}+\mu_{D^{*0}}+\mu_{B^{*+}}+\mu_{B^{*0}}\right),\nonumber\\
\mu_{D^*B^*(2^+)}&=&\frac{1}{2}\left(\mu_{D^{*+}}+\mu_{D^{*0}}+\mu_{B^{*+}}+\mu_{B^{*0}}\right),\nonumber\\
\mu_{D^*_sB^*_s(0^+)}&=&0,\nonumber\\
\mu_{D^*_sB^*_s(1^+)}&=&\frac{1}{2}\left(\mu_{D_s^{*+}}+\mu_{B_s^{*0}}\right),\nonumber\\
\mu_{D^*_sB^*_s(2^+)}&=&\mu_{D_s^{*+}}+\mu_{B_s^{*0}}.
\end{eqnarray}
Based on the above results, the magnetic moments of the $S$-wave isoscalar $B_c$-like molecular states are the combination of the magnetic moments of their constituent hadrons. Thus, the magnetic moments of the $D^{(*)}_{(s)}$ and $B^{(*)}_{(s)}$ mesons are the important input parameters for the study of the magnetic moments of the $S$-wave isoscalar $B_c$-like molecular states.

Based on the numerical results of the magnetic moments of the $D^{(*)}_{(s)}$ and $B^{(*)}_{(s)}$ mesons, we can further obtain the numerical results of the magnetic moments of the $S$-wave isoscalar $B_c$-like molecular states. In Table \ref{TableMBclikeS}, the numerical results of the magnetic moments of the $S$-wave isoscalar $B_c$-like molecular states are collected.

\renewcommand\tabcolsep{0.80cm}
\renewcommand{\arraystretch}{1.50}
\begin{table}[!htbp]
  \caption{The magnetic moments of the $S$-wave isoscalar $B_c$-like molecular states when only considering the $S$-wave component. Here, the magnetic moment of the hadron is in unit of the nuclear magneton $\mu_N$.}
  \label{TableMBclikeS}
\begin{tabular}{cc|c}
\toprule[1.0pt]\toprule[1.0pt]
Systems& $J^P$ &Magnetic moments\\\hline
$DB$& $0^+$ &0\\
$DB^*$& $1^+$ &0.532\\
$D^*B$& $1^+$ &$-0.089$\\
\multirow{3}{*}{$D^*B^*$}& $0^+$ &0\\
                         & $1^+$ &$0.222$\\
                         & $2^+$ &$0.443$\\
\multirow{3}{*}{$D_s^*B_s^*$}& $0^+$ &0\\
                         & $1^+$ &$0.221$\\
                         & $2^+$ &$0.443$\\
\bottomrule[1.0pt]\bottomrule[1.0pt]
\end{tabular}
\end{table}

From the results of the magnetic moments of the $S$-wave isoscalar $B_c$-like molecular states presented in Table \ref{TableMBclikeS}, we can find several interesting results:
\begin{itemize}
  \item The magnetic moments of the $S$-wave $DB$ state with $I(J^P)=0(0^+)$, the $S$-wave $D^*B^*$ state with $I(J^P)=0(0^+)$, and the $S$-wave $D_s^*B_s^*$ state with $I(J^P)=0(0^+)$ are zero, and this is easy to understand since the spin quantum numbers of these $S$-wave isoscalar $B_c$-like molecular states are zero.
  \item The $S$-wave $DB^*$ state with $I(J^P)=0(1^+)$ and the $S$-wave $D^*B$ state with $I(J^P)=0(1^+)$ have the same quantum numbers and quark configurations, but their magnetic moments have obvious differences, which is because the $D^{*}$ and $B^{*}$ mesons have different magnetic moments.
  \item The $S$-wave $D^*B^*$ states and the $S$-wave $D_s^*B_s^*$ states with the same quantum numbers have extremely similar magnetic moments, and their magnetic moments satisfy the relation $\frac{\mu_{D^*B^*(2^+)}}{\mu_{D^*B^*(1^+)}}=\frac{\mu_{D^*_sB^*_s(2^+)}}{\mu_{D^*_sB^*_s(1^+)}}=2$, which can be viewed as the important relation to test our theoretical results by future experiments and other approaches.
\end{itemize}

Then, we further discuss the magnetic moments of the $B_c$-like molecular states after adding the contribution of the $D$-wave channels. For the $D^{*}B^{*}$ and $D^{*}_{s}B^{*}_{s}$ systems of the $D$-wave channels, their magnetic moments can be calculated by the following matrix element \cite{Cheng:1997kr,Liu:2003ab,Huang:2004tn,Haghpayma:2006hu,Sharma:2010vv,Sharma:2012jqz,Girdhar:2015gsa,Wang:2016dzu,Dahiya:2018ahb,Gao:2021hmv,Li:2021ryu,Zhou:2022gra,Wang:2022tib,Wang:2022ugk}:
\begin{eqnarray}
\mu_{H}&=&\left\langle J_{H},J_{H} \left|\sum_{j}\hat{\mu}_{jz}^{\rm spin}+\hat{\mu}_{z}^{\rm orbital} \right| J_{H},J_{H} \right\rangle.
\label{MBclikeSD}
\end{eqnarray}
For the $D^{*}B^{*}$ and $D^{*}_{s}B^{*}_{s}$ systems, the notations $\alpha$ and $\beta$ denote the $D^{*}_{(s)}$ and $B^{*}_{(s)}$ mesons, and $\hat{L}_z$ is the $z$-component of the orbital angular momenta operator between the $D^{*}_{(s)}$ and $B^{*}_{(s)}$ mesons.

In Table \ref{TableMBcSDResults}, we list the numerical results of the magnetic moments of the isoscalar $D^{*}B^{*}$ and $D^{*}_{s}B^{*}_{s}$ molecular states after including the $S$-$D$ wave mixing effect. Here, we also take three typical binding energies $-2$, $-7$, and $-12~{\rm MeV}$ for the initial and final $B_c$-like molecular states to discuss these magnetic moments.

\renewcommand\tabcolsep{0.20cm}
\renewcommand{\arraystretch}{1.50}
\begin{table}[!htbp]
  \caption{The magnetic moments of the isoscalar $D^{*}B^{*}$ and $D^{*}_{s}B^{*}_{s}$ molecules after including the $S$-$D$ wave mixing effect. Here, the magnetic moment of the hadron is in unit of the nuclear magneton $\mu_N$.}
  \label{TableMBcSDResults}
\begin{tabular}{cc|ccc}
\toprule[1.0pt]\toprule[1.0pt]
\multirow{2}{*}{Systems}& \multirow{2}{*}{$J^P$} &\multicolumn{3}{c}{Magnetic moments}\\
                        &                           &$E=-2~{\rm MeV}$&$E=-7~{\rm MeV}$&$E=-12~{\rm MeV}$\\\hline
\multirow{3}{*}{$D^*B^*$}& $0^+$ &0            &0                 &0\\
                         & $1^+$ &$0.219$      &$0.218$           &$0.217$\\
                         & $2^+$ &$0.442$      &$0.441$           &$0.441$\\
\multirow{3}{*}{$D_s^*B_s^*$}& $0^+$ &0            &0                 &0\\
                         & $1^+$ &$0.225$          &$0.227$           &$0.228$\\
                         & $2^+$ &$0.446$          &$0.449$           &$0.450$\\
\bottomrule[1.0pt]\bottomrule[1.0pt]
\end{tabular}
\end{table}

The magnetic moments of the $|^{1} S_{0}\rangle$ and $|^{5} D_{0}\rangle$ channels for the isoscalar $D^{*}B^{*}$ and $D^{*}_{s}B^{*}_{s}$ systems are zero, which implies that the magnetic moments of the isoscalar $D^*B^*$ and $D_s^*B_s^*$ molecules with $J^P=0^+$ are still zero after considering the $S$-$D$ wave mixing effect. By comparing the obtained results of the single channel and $S$-$D$ wave mixing analysis, the magnetic moments of the isoscalar $D^{*}B^{*}$ and $D^{*}_{s}B^{*}_{s}$ molecules with $J^P=1^+$ and $2^+$ will change accordingly after considering the $S$-$D$ wave mixing effect. However, the change of their magnetic moments is less than $0.07\mu_N$, and the significant change are the isoscalar $D_s^*B_s^*$ molecules with $J^P=1^+$ and $2^+$. Thus, the $S$-$D$ wave mixing effect plays a minor role to modify the magnetic moments of the isoscalar $D^{*}B^{*}$ and $D^{*}_{s}B^{*}_{s}$ molecules.

\subsection{Difference of the magnetic moments of the $B_c$-like molecule, the compact $B_c$-like tetraquark, and the $B_c$ mesonic state}

In this subsection, we want to answer whether or not the magnetic moment properties can be used to distinguish the compact $B_c$-like tetraquark states and the $B_c$-like molecular states, or the conventional $B_c$ mesonic states and the $B_c$-like molecular states, which may provide crucial information to establish the mass spectra of these hadronic states.

\begin{figure}[!htbp]
  \centering
  \includegraphics[width=0.40\textwidth]{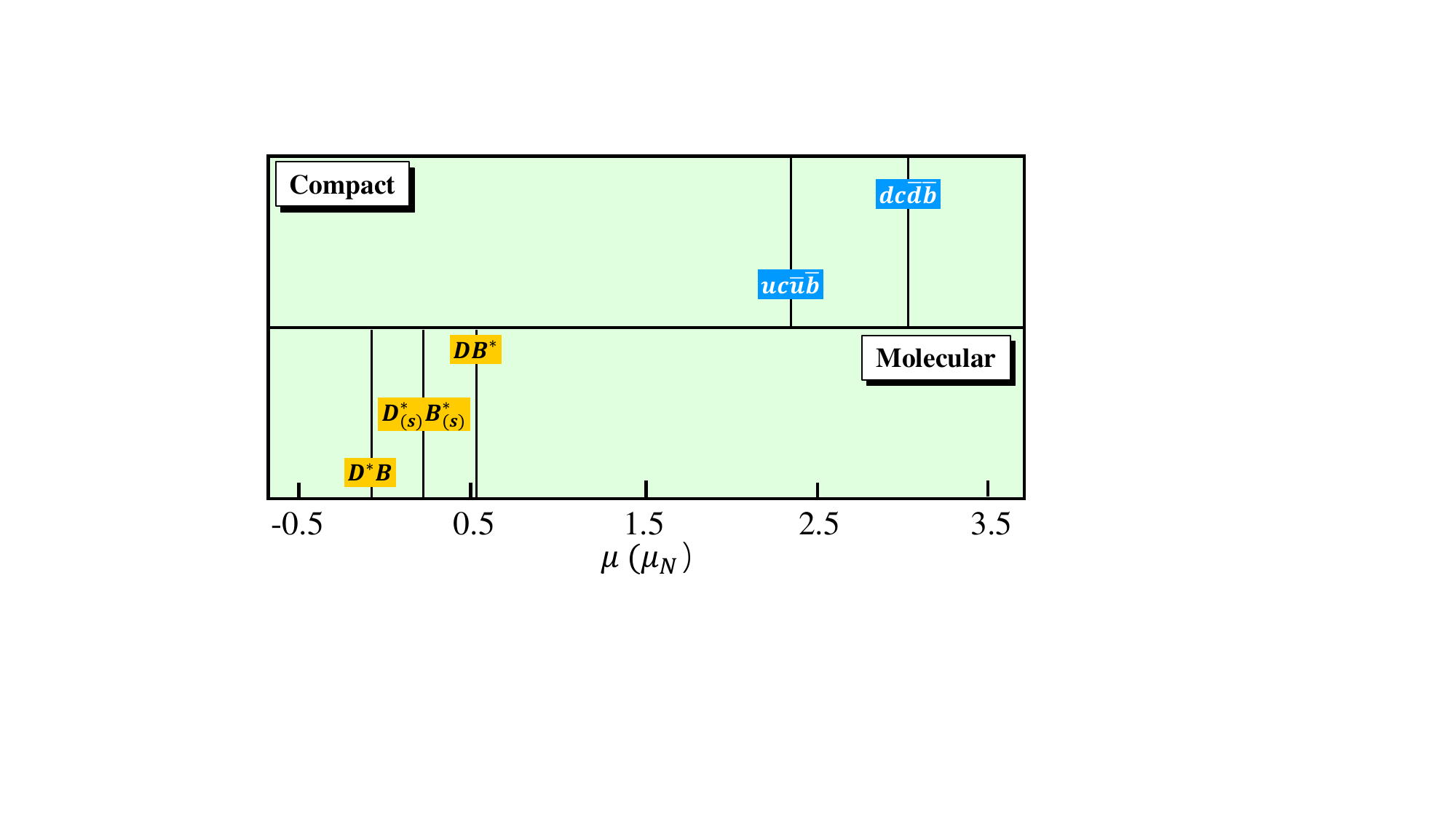}
  \caption{The comparison of the magnetic moments of the compact $B_c$-like tetraquark states \cite{Ozdem:2022eds} and the $B_c$-like molecular states with $I(J^P)=0(1^+)$. Here, the components corresponding to the concrete states are marked. }\label{Comparasion}
\end{figure}

In Ref. \cite{Ozdem:2022eds}, the author already discussed the magnetic moments of the compact $B_c$-like tetraquark states with $I(J^P)=0(1^+)$ in the diquark-antidiquark picture within the QCD light-cone sum rules. In Fig. \ref{Comparasion}, we compare the magnetic moments of the compact $B_c$-like tetraquark states \cite{Ozdem:2022eds} and the $B_c$-like molecular states with $I(J^P)=0(1^+)$. As shown in Fig. \ref{Comparasion}, the magnetic moments of the compact $B_c$-like tetraquark states with $I(J^P)=0(1^+)$ are larger than $2.30{\rm \mu_N}$ \cite{Ozdem:2022eds}, while the magnetic moments of the $B_c$-like molecular states with $I(J^P)=0(1^+)$ are smaller than $0.60{\rm \mu_N}$. Thus, the compact $B_c$-like tetraquark states and the $B_c$-like molecular states with $I(J^P)=0(1^+)$ have significantly different magnetic moment properties, which shows that the measurement of the magnetic moment properties can be used to distinguish the compact $B_c$-like tetraquark states and the $B_c$-like molecular states in the future experiments.

By comparing the mass spectra of the conventional $B_c$ mesonic states \cite{Godfrey:1985xj,Eichten:1994gt,Gershtein:1994dxw,Zeng:1994vj,Ebert:2002pp,Godfrey:2004ya,Soni:2017wvy,Eichten:2019gig,Li:2019tbn,Li:2022bre} and the predicted $B_c$-like molecular states \cite{Sun:2012sy}, some conventional $B_c$ mesonic states and some $B_c$-like molecular states have the same quantum numbers and similar masses. For example, the $B_c(2P_1^{\prime})$ state and the $DB^*$ state with $I(J^P)=0(1^+)$, the $B_c(2P_1)$ state and the $DB^*$ state with $I(J^P)=0(1^+)$, the $B_c(3P_1^{\prime})$ state and the $D_s^*B_s^*$ state with $I(J^P)=0(1^+)$, the $B_c(3{}^3P_2)$ state and the $D_s^*B_s^*$ state with $I(J^P)=0(2^+)$, and so on. Thus, there may be a great challenge when establishing the mass spectra of the conventional $B_c$ mesonic states and the $B_c$-like molecular states. Facing this situation mentioned above, we need to answer whether or not the magnetic moment properties can be used to distinguish the conventional $B_c$ mesonic states and the $B_c$-like molecular states with the same quantum numbers and similar masses.

In the following, we discuss the magnetic moments of the $B_c(2P_1^{\prime})$, $B_c(2P_1)$, $B_c(3P_1^{\prime})$, and $B_c(3{}^3P_2)$ states within the constituent quark model. For the conventional $B_c$ mesonic states, we need to specify that the notations $\alpha$ and $\beta$ denote the $c$ and $\bar b$ quarks, and $\hat{L}_z$ is the $z$-component of the orbital angular momenta operator between the $c$ and $\bar b$ quarks in Eq. (\ref{MBclikeSD}). After expanding the spin-orbital wave functions $|{ }^{2 S+1} L_{J}\rangle$ by the orbital wave function $Y_{L, m_{L}}$ and the spin wave function $\chi_{S,\,m_{S}}$, we can get the following relations:
\begin{eqnarray}
\left|{ }^{1} P_{1}\right\rangle&=&Y_{1, 1}\chi_{0,0},\nonumber\\
\left|{ }^{3} P_{1}\right\rangle&=&\frac{1}{\sqrt{2}}Y_{1, 1}\chi_{1,0}-\frac{1}{\sqrt{2}}Y_{1, 0}\chi_{1,1},\nonumber\\
\left|{ }^{3} P_{2}\right\rangle&=&Y_{1, 1}\chi_{1,1}.
\end{eqnarray}
Thus, the magnetic moments and the transition magnetic moment of the conventional $B_c$ mesonic states are
\begin{eqnarray}
\mu_{B_c\left|{ }^{1} P_{1}\right\rangle}&=&\mu_{c \bar b}^L=0.296~{\rm \mu_N},\nonumber\\
\mu_{B_c\left|{ }^{3} P_{1}\right\rangle}&=&\frac{1}{2}\mu_{c}+\frac{1}{2}\mu_{\bar b}+\frac{1}{2}\mu_{c \bar b}^L=0.370~{\rm \mu_N},\nonumber\\
\mu_{B_c\left|{ }^{3} P_{1}\right\rangle \to B_c\left|{ }^{1} P_{1}\right\rangle}&=&\frac{1}{\sqrt 2}\mu_{c}-\frac{1}{\sqrt 2}\mu_{\bar b}=0.220~{\rm \mu_N},\nonumber\\
\mu_{B_c\left|{ }^{3} P_{2}\right\rangle}&=&\mu_{c}+\mu_{\bar b}+\mu_{c \bar b}^L=0.739~{\rm \mu_N}.
\end{eqnarray}
In the above expressions, we use the constituent quark masses $m_c=1.660~{\rm GeV}$ and $m_b=4.730~{\rm GeV}$ \cite{Lichtenberg:1976fi,Li:2017cfz,Meng:2017dni,Li:2017pxa,Wang:2019mhm,Gao:2021hmv} to present the magnetic moments and the transition magnetic moment of the conventional $B_c$ mesonic states. Furthermore, the $P$-wave physical states of the conventional $B_c$ mesonic states are the linear combination of the $\left|{}^1P_1\right\rangle$ and $\left|{}^3P_1\right\rangle$ states, which can be described by \cite{Eichten:2019gig}
\begin{eqnarray}
\left|nP_1^{\prime}\right\rangle&=&\left|n{}^1P_1\right\rangle{\rm cos}\theta_{nP}+\left|n{}^3P_1\right\rangle{\rm sin}\theta_{nP},\nonumber\\
\left|nP_1\right\rangle&=&-\left|n{}^1P_1\right\rangle{\rm sin}\theta_{nP}+\left|n{}^3P_1\right\rangle{\rm cos}\theta_{nP}.
\end{eqnarray}
Here, the related mixing angles are $\theta_{2P}=18.7^{\rm o}$ and $\theta_{3P}=21.2^{\rm o}$ \cite{Eichten:2019gig}. According to the above preparation, the magnetic moments of the $B_c(2P_1^{\prime})$, $B_c(2P_1)$, $B_c(3P_1^{\prime})$, and $B_c(3{}^3P_2)$ states can be written as
\begin{eqnarray}
\mu_{B_c(2P_1^{\prime})}&=&\mu_{\left|{ }^{1} P_{1}\right\rangle}{\rm cos}^2\theta_{2P}+\mu_{\left|{ }^{3} P_{1}\right\rangle \to \left|{ }^{1} P_{1}\right\rangle}{\rm sin}2\theta_{2P}+\mu_{\left|{ }^{3} P_{1}\right\rangle}{\rm sin}^2\theta_{2P}\nonumber\\
&=&0.437~{\rm \mu_N},\nonumber\\
%%%%%%%%%%%%%%%%%%%%%%%
\mu_{B_c(2P_1)}&=&\mu_{\left|{ }^{1} P_{1}\right\rangle}{\rm sin}^2\theta_{2P}-\mu_{\left|{ }^{3} P_{1}\right\rangle \to \left|{ }^{1} P_{1}\right\rangle}{\rm sin}2\theta_{2P}+\mu_{\left|{ }^{3} P_{1}\right\rangle}{\rm cos}^2\theta_{2P}\nonumber\\
&=&0.229~{\rm \mu_N},\nonumber\\
%%%%%%%%%%%%%
\mu_{B_c(3P_1^{\prime})}&=&\mu_{\left|{ }^{1} P_{1}\right\rangle}{\rm cos}^2\theta_{3P}+\mu_{\left|{ }^{3} P_{1}\right\rangle \to \left|{ }^{1} P_{1}\right\rangle}{\rm sin}2\theta_{3P}+\mu_{\left|{ }^{3} P_{1}\right\rangle}{\rm sin}^2\theta_{3P}\nonumber\\
&=&0.454~{\rm \mu_N},\nonumber\\
%%%%%%%%%%%%%
\mu_{B_c(3{}^3P_2)}&=&0.739~{\rm \mu_N},
\end{eqnarray}
respectively. When studying the magnetic moments of the $B_c(2P_1^{\prime})$, $B_c(2P_1)$, and $B_c(3P_1^{\prime})$ states, we ignore the contribution of the factor $\left\langle R_{B_c^{\prime}} \left|e^{-i {\bf k}\cdot{\bf r}_j}\right| R_{B_c}\right\rangle$ for the transition magnetic moments of the $|2{ }^{3} P_{1}\rangle\to|2{ }^{1} P_{1}\rangle \gamma$ and $|3{ }^{3} P_{1}\rangle\to|3{ }^{1} P_{1}\rangle \gamma$ processes, since the momenta of the emitted photon are less than $4~{\rm MeV}$ for these radiative decay processes \cite{Soni:2017wvy}.

In Table \ref{TablecomparisonMT}, we compare the magnetic moments of several $B_c$-like molecular states and conventional $B_c$ mesonic states with the same quantum numbers and similar masses, and we find that the magnetic moments of these $B_c$-like molecular states and conventional $B_c$ mesonic states have differences. In short, the magnetic moment properties can provide the crucial information to distinguish the $B_c$-like molecular states and the conventional $B_c$ mesonic states with the same quantum numbers and similar masses in the future experiments, and we wish that more theoretical and experimental colleagues can pay more attention to focus on the magnetic moment properties of the $B_c$-like molecular states and the conventional $B_c$ mesonic states, which can provide more abundant suggestions to identify the $B_c$-like molecular states and the conventional $B_c$ mesonic states with the same quantum numbers and similar masses.
\renewcommand\tabcolsep{0.33cm}
\renewcommand{\arraystretch}{1.50}
\begin{table}[!htbp]
  \caption{The comparison of the magnetic moments of several $B_c$-like molecular states and conventional $B_c$ mesonic states with the same quantum numbers and similar masses. Here, the $B_c$-like molecular states and the conventional $B_c$ mesonic states in the same row have the same quantum numbers and similar masses.}
  \label{TablecomparisonMT}
\begin{tabular}{cc|cc}
\toprule[1.0pt]\toprule[1.0pt]
\multicolumn{2}{c|}{$B_c$-like molecular states} & \multicolumn{2}{c}{Conventional $B_c$ mesonic states} \\\hline
$DB^*[0(1^+)]$ & $0.532~{\rm \mu_N}$& $B_c(2P_1^{\prime})$&$0.437~{\rm \mu_N}$\\
$DB^*[0(1^+)]$ & $0.532~{\rm \mu_N}$& $B_c(2P_1)$ &$0.229~{\rm \mu_N}$\\
$D_s^*B_s^*[0(1^+)]$ & $0.221~{\rm \mu_N}$& $B_c(3P_1^{\prime})$&$0.454~{\rm \mu_N}$\\
$D_s^*B_s^*[0(2^+)]$ & $0.443~{\rm \mu_N}$&$B_c(3{}^3P_2)$&$0.739~{\rm \mu_N}$\\
\bottomrule[1.0pt]
\bottomrule[1.0pt]
\end{tabular}
\end{table}

Based on the above analysis, we can conclude that the magnetic moment properties can be considered as the effective physical observable to distinguish the compact $B_c$-like tetraquark states and the $B_c$-like molecular states, or the conventional $B_c$ mesonic states and the $B_c$-like molecular states. Therefore, we hope that the future experiments can bring us more surprises when analyzing the magnetic moment properties of the conventional $B_c$ mesonic states, the $B_c$-like molecular states, and the compact $B_c$-like tetraquark states.

Before closing this section, we need to specify that the magnetic moments of the unstable particles can be the complex-valued quantities. For example, the $\Delta(1232)$ \cite{Anderson:1952nw} is the unstable state, and its magnetic moment can be the complex-valued quantity \cite{Pascalutsa:2004je,Pascalutsa:2007wb,Pascalutsa:2006up}, which is due to the intermediate pion-nucleon state being able to go on shell, generating the imaginary part for its magnetic moment. However, such behavior is not covered in the constituent quark model \cite{Schlumpf:1993rm,Kumar:2005ei}. For these discussed $B_c$-like molecular states, they are the unstable particles, and their magnetic moments can be the complex-valued quantities. In the present work, we only give the real parts for their magnetic moments by adopting the constituent quark model. In addition, we hope that the magnetic moments of the $B_c$-like molecular states can be discussed by other models and approaches in the future, and these investigations will make our knowledge of the magnetic moments of the $B_c$-like molecular states become more abundant.

\section{Summary}\label{sec4}

As an important and active research topic in hadron spectroscopy, the exploration of the hadronic molecular states is full of opportunities and challenges. Since the discovery of the charmonium-like state $X(3872)$ in 2003, there have been extensive experimental and theoretical investigations around the hadronic molecules, which is because the masses of a number of new hadronic states are very close to the corresponding thresholds of two hadrons. In addition to providing their mass spectra, we still need to make more efforts to reveal other aspects of the hadronic molecular states, where their electromagnetic properties are the valuable physical observable. The electromagnetic properties of the hadronic molecular states mainly include the radiative decay widths and the magnetic moments, which may reflect their inner structures.

In Ref. \cite{Sun:2012sy}, the mass spectra of the $B_c$-like molecular states composed of the $D^{(*)}_{(s)}$ and $B^{(*)}_{(s)}$ mesons were investigated. For further disclosing the properties of the $B_c$-like molecular states, in this work we first study the transition magnetic moments and the corresponding radiative decay widths of the $B_c$-like molecular states associated with their mass spectra, where the constituent quark model is applied. In the concrete calculation, we consider the contribution of the spatial wave functions of the initial and final states. Our numerical results show that there exist different radiative decay widths for these discussed  $B_c$-like molecular states, which strongly depend on the transition magnetic moments and the phase spaces, and several radiative decay processes of the $B_c$-like molecular states have significant widths. In addition, we also discuss the role of the $S$-$D$ wave mixing effect to the transition magnetic moments of the isoscalar $D^{*}B^{*}$ and $D^{*}_{s}B^{*}_{s}$ molecules, and we find the $S$-$D$ wave mixing effect plays a minor role to decorate the transition magnetic moments and the radiative decay widths of the isoscalar $D^{*}B^{*}$ and $D^{*}_{s}B^{*}_{s}$ molecular states.

Meanwhile, we also discuss the magnetic moments of the $B_c$-like molecular states. By performing a quantitative calculation, we find that (i) the $D^*B^*$ molecules and the $D_s^*B_s^*$ molecules with the same quantum numbers have extremely similar magnetic moments, (ii) the $DB^*$ state with $I(J^P)=0(1^+)$ and the $D^*B$ state with $I(J^P)=0(1^+)$ have obviously different magnetic moments, and (iii) the magnetic moment properties can be considered as the effective physical observable to distinguish the hadrons with different configurations, especially with identifying the conventional $B_c$ mesonic states and the $B_c$-like molecular states with the same quantum numbers and similar masses.

Exploration of the electromagnetic properties of the hadronic molecular states may provide new insights to reflect their inner structures. In particular, the magnetic moment properties can provide the important physical observable to distinguish the hadrons with different configurations or spin-parity quantum numbers. As a potential research topic full of opportunities and challenges, exploring the electromagnetic properties of the hadronic molecular states should be given more attention by both theorist and experimentalist, which can make our knowledge of the hadronic molecules become more complete.

\section*{Acknowledgement}

This work is supported by the China National Funds for Distinguished Young Scientists under Grant No. 11825503, the National Key Research and Development Program of China under Contract No. 2020YFA0406400, the 111 Project under Grant No. B20063, the National Natural Science Foundation of China under Grant Nos. 12247101 and 12247155, and the project for top-notch innovative talents of Gansu province.
F.L.W is also supported by the China Postdoctoral Science Foundation under Grant No. 2022M721440.

\end{document}